\documentclass[amssymb,nofootinbib,nobibnotes,aps,prd,preprintnumbers]{revtex4}

 \usepackage{here}

\usepackage[dvipdfmx]{graphicx}
\usepackage{amssymb,amsfonts,amsmath,bm,color,multirow,cases,empheq,hyperref,ascmac,fancybox}
\usepackage{mathrsfs}

\definecolor{dgreen}{rgb}{0.0, 0.5, 0.0}

\usepackage{enumerate}
\usepackage{ulem}
\usepackage{afterpage}
\usepackage{amsmath}

\hypersetup{%
 setpagesize=false,
 bookmarksnumbered=true,%
 bookmarksopen=true,%
 colorlinks=true,%
 linkcolor=blue,%
 citecolor=red}

\begin{document}




\title{Fan--Wang type regular black holes in Quasi-Topological Gravity}

\author{Ren Tsuda}
\email{tsuda.ren@chibatech.ac.jp}
\affiliation{Student Support Center, Chiba Institute of Technology\\
Shibazono 2-1-1, Narashino 275-0023, Japan}

\author{Ryotaku Suzuki}
\email{suzuki.ryotaku@nihon-u.ac.jp}
\affiliation{Laboratory of Physics, College of Science and Technology, Nihon University\\
Narashinodai 7-24-1, Funabashi 274-8501, Japan}

\author{Shinya Tomizawa}
\email{tomizawa@toyota-ti.ac.jp}
\affiliation{Mathematical Physics Laboratory, Toyota Technological Institute\\
Hisakata 2-12-1, Nagoya 468-8511, Japan}

\date{\today}

\preprint{TTI-MATHPHYS-39}




\begin{abstract}

We construct a class of regular black hole solutions of the Fan--Wang type within quasi-topological gravity (QTG) in arbitrary spacetime dimensions greater than four. In contrast to the original Fan--Wang solution, which was obtained in four-dimensional general relativity coupled to nonlinear electrodynamics, our higher-dimensional generalization does not require any matter fields. Instead, regularity is achieved purely through an infinite tower of higher-curvature corrections.
We demonstrate that the Fan--Wang-type metric is a solution to the QTG field equations by explicitly determining the corresponding coupling constants for each curvature order. Within an appropriate parameter regime, the solution describes an asymptotically flat black hole spacetime with a regular center. Remarkably, even in the case of negative mass, the geometry can remain completely regular, in sharp contrast to Einstein gravity.

\end{abstract}

\maketitle




\section{Introduction}

\label{sec:intro}

The existence of black holes is now firmly established through a wide range of observational breakthroughs in gravitational physics and high-energy astrophysics.
In particular, the first direct detection of gravitational waves from binary black hole mergers by the LIGO/Virgo collaboration~\cite{LIGOScientific:2016aoc} has provided compelling dynamical evidence for the existence of compact objects whose properties are in remarkable agreement with the predictions of general relativity.
More recently, the Event Horizon Telescope has directly imaged the shadow of supermassive black holes at the centers of galaxies~\cite{EventHorizonTelescope:2019dse}, offering an unprecedented probe of strong-field gravity near the event horizon.
These achievements not only confirm the astrophysical reality of black holes but also open a new observational window onto the nonlinear regime of gravitational dynamics.
At the same time, black holes continue to play a central role in theoretical investigations ranging from classical gravity and quantum field theory in curved spacetime to holography and quantum information.
However, within classical general relativity, physically reasonable energy conditions generically imply the formation of spacetime singularities inside black holes, as exemplified by Penrose's singularity theorem~\cite{Penrose:1964wq}.
Such singularities signal a breakdown of the classical description, where curvature invariants diverge and geodesic incompleteness arises, indicating that the theory ceases to provide predictive power at sufficiently small length scales.
It is widely expected that these singularities are artifacts of the classical approximation and that they should be resolved once genuine quantum-gravity effects, or an appropriate ultraviolet completion of gravity, are consistently incorporated.
In this sense, black hole interiors provide a natural arena in which the interplay between strong curvature and quantum effects becomes unavoidable.
In this context, the construction and analysis of singularity-free black hole spacetimes, usually referred to as regular black holes, can offer a valuable phenomenological window into the expected short-distance modifications of gravity and may shed light on possible mechanisms of singularity resolution beyond Einstein theory.

\medskip

A regular black hole is typically defined as a spacetime that possesses an event horizon while all curvature invariants remain finite everywhere, including the black hole interior.
In other words, the geometry is free from curvature singularities and geodesically complete (or at least free of incomplete geodesics associated with divergent curvature), despite exhibiting horizon structure analogous to that of ordinary black holes.
The first celebrated example is the Bardeen model~\cite{Bardeen:1968aa}, originally proposed as an effective metric ansatz rather than as a solution of a known field theory.
Its line element interpolates smoothly between an asymptotically Schwarzschild-like exterior and a de Sitter-like core near the center, thereby avoiding the $r=0$ singularity of the Schwarzschild solution.
Subsequently, various regular black hole models were introduced, for instance by Dymnikova and Hayward~\cite{Dymnikova:1992ux,Hayward:2005gi}, and their general structural properties --- such as the emergence of an effective de Sitter core, the violation of the strong energy condition near the center, and the existence of multiple horizons depending on parameters --- have been analyzed in detail; see, e.g., the review~\cite{Ansoldi:2008jw}.
A major step toward embedding these geometries into a consistent field-theoretic framework was achieved by Ay{\'o}n-Beato and Garc{\'i}a, who demonstrated that Bardeen-type regular black holes can arise as exact solutions of Einstein gravity coupled to nonlinear electrodynamics (NED)~\cite{Ayon-Beato:1998hmi,Ayon-Beato:2000mjt}.
In this construction, the effective matter sector supporting the regular core is provided by a suitably engineered nonlinear electromagnetic Lagrangian, which modifies the stress-energy tensor in the high-field regime while preserving standard Maxwell behavior asymptotically.
More generally, Fan and Wang constructed a broad class of static, spherically symmetric regular black holes (including the Bardeen and Hayward solutions as special cases) within Einstein--NED systems~\cite{Fan:2016hvf}, clarifying how different choices of mass functions and NED Lagrangians give rise to distinct horizon structures and core behaviors.
Many further examples have since been developed along similar lines~\cite{Dymnikova:2015hka,Takeuchi:2016nrj,Kruglov:2017fck,Maeda:2021jdc,Bokulic:2022cyk,Bronnikov:2022ofk}, exploring  thermodynamic aspects of regular geometries.
Beyond NED-based models, regular black hole solutions have also been investigated in a variety of alternative frameworks, including effective quantum-gravity inspired scenarios (such as loop-inspired metrics), nonlocal gravity models, and modified gravity theories with higher-curvature corrections.
In these settings, additional gravitational degrees of freedom or curvature corrections can alter the short-distance structure of the field equations and naturally generate nonsingular cores without introducing ad hoc matter sources; for complementary discussions, see also~\cite{Frolov:2016pav}.
Together, these developments suggest that regular black holes provide a useful bridge between phenomenological modifications of gravity and more fundamental approaches to quantum gravity.

\medskip

Among modified-gravity frameworks, higher-curvature theories are particularly well-motivated because curvature corrections naturally arise in low-energy effective actions of quantum gravity (e.g., string-inspired expansions).
While generic higher-derivative theories typically lead to higher-order equations of motion, there exist distinguished classes in which the dynamics remains tractable for highly symmetric ans\"atze.
A prominent example is Lovelock gravity, whose field equations are second order in any dimension, and which contains the Gauss--Bonnet term as the leading nontrivial correction in $D\ge5$.
In five dimensions and higher, another important family is provided by {\it quasi-topological gravity} (QTG), in which specific higher-curvature combinations (cubic and beyond) are engineered so that, for static and spherically symmetric spacetimes, the equations reduce to a second-order form while retaining a healthy linearized graviton spectrum around maximally symmetric backgrounds \cite{Oliva:2010eb,Myers:2010ru,Dehghani:2011vu, Cisterna:2017umf}.
This feature allows one to analyze black hole solutions and their thermodynamics with a level of analytic control comparable to Einstein and Lovelock theories.
More recently, these ideas have been extended to broader classes, including generalized quasi-topological gravities and closely related ``Einsteinian'' higher-curvature theories, which share similar simplifications in symmetric sectors and have been actively studied from both gravitational and holographic viewpoints \cite{Bueno:2016xff,Hennigar:2017ego,Bueno:2019ycr, Aguayo:2025xfi}.
It is therefore natural to ask whether such higher-curvature corrections can also accommodate regular black holes, either by themselves or in combination with suitable matter sectors.
Indeed, concrete constructions of regular black holes sourced by quasi-topological gravities in $D\ge5$ have appeared recently~\cite{Fernandes:2025fnz,Bueno:2024dgm}, providing new examples in which nonsingular interiors arise within a higher-curvature setting. This type of regular black hole has recently been constructed in the presence of Maxwell field~\cite{Hao:2025utc}.
These developments further motivate a systematic investigation of regular black hole solutions in higher dimensions and in gravity theories beyond Einstein's.

\medskip
In this paper we construct and analyze a new family of higher-dimensional regular black holes that can be viewed as the vacuum, quasi-topological-gravity (QTG) counterpart of the four-dimensional Fan--Wang solution originally obtained in Einstein gravity coupled to nonlinear electrodynamics (NED)~\cite{Fan:2016hvf}. 
The basic viewpoint is that the ``effective matter'' responsible for regularizing the central region in NED models can, in higher dimensions, be mimicked by an infinite tower of higher-curvature corrections in the gravitational action. 
Working within QTG, where static and spherically symmetric field equations reduce to an algebraic relation between the metric function and a polynomial-like function $h(\psi)$, we show that one can engineer an $h(\psi)$ (equivalently, an infinite set of QTG couplings) so that the resulting vacuum solution reproduces a Fan--Wang--type metric profile in $D>4$. 
In practice, this requires going beyond finitely many curvature terms: regularity at the center forces a resummed/infinite-curvature structure, and we determine the corresponding couplings systematically by reconstructing $h(\psi)$ from the desired metric function.
A key result is that demanding a nonsingular center tightly constrains the parameter space of the would-be Fan--Wang generalization. 
In particular, the regularity condition at $r=0$
together with consistency of the QTG relation uniquely fixes the relevant parameter of the theory to a specific value (in our notation, $\mu=3$). 
With this choice, the geometry develops a de Sitter-like core and all curvature invariants remain finite at the center, while the spacetime remains asymptotically flat at infinity. 
We further clarify the smoothness properties of the geometry at $r=0$ in Cartesian coordinates: depending on the dimension and discrete parameters in the construction, the spacetime can be fully smooth or only finitely differentiable, yet still curvature-regular. 
We then investigate the global and thermodynamic properties of the resulting solutions. 
We  classify the parameter ranges that yield two-horizon, extremal, or horizonless configurations, and we show that the event horizon(s) are regular by exhibiting coordinate systems that extend smoothly across them. 
Thermodynamics is studied within QTG using the appropriate higher-curvature notions of conserved mass and Wald entropy, leading to explicit expressions for temperature, entropy, and extremality conditions, and we verify their mutual consistency (including the first-law structure). 
We also comment on less standard regimes--such as solutions with negative mass parameter--where higher-curvature effects can qualitatively alter the usual expectations from Einstein gravity, and  
we clarify the conditions under which the geometry remains regular and those under which pathological behaviors, such as naked singularities, arise.
Overall, our results demonstrate that Fan--Wang--like regular black holes admit a natural and explicit realization in higher-dimensional vacuum QTG, and they provide a concrete framework for studying singularity resolution driven purely by gravitational higher-curvature terms rather than by additional matter.

\medskip 

The remainder of this paper is organized as follows.
In Sec.~\ref{sec:overview}, we briefly review the framework of QTG and explain how an infinite tower of higher-curvature corrections can yield regular black holes as vacuum solutions.
In Sec.~\ref{sec:FWGR}, we review the four-dimensional Fan--Wang black holes in general relativity coupled to nonlinear electrodynamics (NED).
We then construct their higher-dimensional generalization within QTG in Sec.~\ref{sec:hdfw} by fixing the couplings of the higher-curvature terms.
We further analyze the resulting solution, including regularity at the center, asymptotic flatness, thermodynamics, the possibility of negative-mass configurations, horizon structure, and regularity at the horizons.
Finally, in Sec.~\ref{sec:summary}, we summarize our results and discuss possible directions for future work.




\section{Overview of Quasi-topological gravity}
\label{sec:overview}

The gravitational action of the QTG theory in $D$ dimensions is given by
\begin{align}
\label{eq:action}
I_{\rm QT} = \frac{1}{16 \pi G} \int d^D x \sqrt{-g} \left[ R + \sum_{n=2}^{n_{\max}} \alpha_n {\cal Z}_n \right],
\end{align}
where $\alpha_n$ are arbitrary coupling constants with dimensions of length$^{2(n-1)}$ and $ {\cal Z}_n $ are $n$-th polynomials of the Riemann curvature tensor, called {\it quasi-topological densities}
(see Ref.~\cite{Bueno:2024dgm} for 
explicit expressions of ${\cal Z}_n$).
Note that these higher curvature corrections contribute only in the dimensions larger than four as long as we consider polynomial corrections.
In four dimensions, all the corrections become topological terms and the theory reduces to the Einstein gravity. One may introduce similar correction terms in four dimensions if one allows $Z_n$ to be  rational functions of the curvature tensor rather than polynomials~\cite{Bueno:2025zaj}.
In this paper, we focus on the $D>4$ case.

\medskip

We now consider the most general static and spherically symmetric ansatz
\begin{align}
ds^2 = - N (r)^2 f (r)\, dt^2 + \frac{dr^2}{f(r)} + r^2 d \Omega^2_{D-2},
\end{align}
for which the field equations reduce to
\begin{align}
\frac{dN(r)}{dr} =0 , \qquad \frac{d}{dr} \left[ r^{D-1} h (\psi) \right] = 0,
\end{align}
with
\begin{align}
\label{eq:psidef}
h (\psi) := \psi + \sum_{n=2}^{n_{\max}} \alpha_n \psi^n , \qquad
\psi := \frac{ 1 - f(r) }{ r^2 }.
\end{align}
Hence $N(r)$ is a constant and can be fixed by normalization to $N(r)=1$.
Moreover, the second equation integrates immediately to give
\begin{align}
\label{eq:hsolution}
h(\psi) = \frac{m}{ r^{D-1} },
\end{align}
where $m$ is an integration constant.

\medskip

For the moment, we assume that $n_{\max}$ is finite.
Then the highest-order curvature correction dominates near the center, and
Eq.~(\ref{eq:hsolution}) is approximated in the limit $r\to 0$ by
\begin{align}
\alpha_{n_{\max}}\psi^{n_{\max}} \approx \frac{m}{r^{D-1}} .
\end{align}
This immediately yields
\begin{align}
f(r) \approx 1 - \sqrt[n_{\max}]{\frac{m}{\alpha_{n_{\max}}}}\,
r^{\,2-\frac{D-1}{n_{\max}}}\qquad (r\ll 1),
\end{align}
and
\begin{align}
\psi(r) \approx \sqrt[n_{\max}]{\frac{m}{\alpha_{n_{\max}}}}\,
r^{-\frac{D-1}{n_{\max}}}\qquad (r\ll 1).
\end{align}
Since the curvature invariants can be expressed in terms of $\psi$ as
\begin{align}
\label{eq:ricciscal}
R =& \frac{ \left( \psi r^D \right)'' }{r^{D-2}}, \\
\label{eq:riccisquared}
R_{\mu \nu} R^{\mu \nu} 
=& \frac{1}{2} \left( \frac{ \left( r^{D-2} \left( \psi r^2 \right)' \right)' }{ r^{D-2} } \right)^2 
+ (D-2) \left( \frac{ \left( \psi r^{D-1} \right)' }{ r^{D-2} } \right)^2, \\
\label{eq:kretschmann}
R_{\mu \nu \rho \sigma} R^{\mu \nu \rho \sigma} 
=& \left( \left( \psi r^2 \right)'' \right)^2 
+ 2 (D-2) \left( \frac{ \left( \psi r^2 \right)' }{r} \right)^2 
+ 2 (D-2) (D-3) \psi^2,
\end{align}
we see that these quantities diverge at $r=0$ unless
\begin{align}
\label{eq:regularity_f}
1 - f(r) = {\cal O}\!\left(r^2\right)
\qquad {\rm or}\qquad
\psi(r) = {\cal O}(r^0)
\qquad (r\to 0).
\end{align}
This condition can only be satisfied in the limit
\begin{align}
\label{eq:nmax}
n_{\max}\to\infty.
\end{align}
Accordingly, in what follows we assume an infinite tower of higher-curvature corrections in the action~(\ref{eq:action}) and, hence, in the field equations~(\ref{eq:psidef}) of QTG.

\medskip
Following Ref.~\cite{Bueno:2024dgm}, one can obtain various families of regular black holes by appropriately tuning the infinitely many couplings $\alpha_n$.
As a simple illustrative example, consider the geometric choice
\begin{align}
\alpha_n=\alpha^{\,n-1},
\end{align}
where $\alpha>0$ is a constant.
Then the series in $h(\psi)$ can be summed explicitly:
\begin{align}
h(\psi)
&=\psi+\sum_{n=2}^{\infty}\alpha^{\,n-1}\psi^n
=\psi\sum_{k=0}^{\infty}(\alpha\psi)^k
=\frac{\psi}{1-\alpha\psi}, 
\end{align}
where the convergence radius is given by $|\psi| < \alpha^{-1}$.
Substituting this expression into Eq.~(\ref{eq:hsolution}), we can determine $\psi$ as a function of $r$.
Then, using the definition $\psi=(1-f(r))/r^2$ in Eq.~(\ref{eq:psidef}), we obtain the metric function
\begin{align}
\label{eq:Hayward}
f(r)=1-\frac{m r^2}{r^{D-1}+\alpha m}.
\end{align}
As studied in Ref.~\cite{Bueno:2024dgm}, this solution can be interpreted as the higher-dimensional analogue of the Hayward regular black hole~\cite{Hayward:2005gi}.




\section{Fan--Wang black holes in four dimensions}\label{sec:FWGR}

Fan and Wang~\cite{Fan:2016hvf} showed that the following class of spacetimes
\begin{align}
\label{eq:fw}
ds^2 = - f(r)\,dt^2 + \frac{dr^2}{ f(r) } + r^2 d \Omega_2^2,
\qquad
f(r) = 1 - \frac{2 M}{r} - \frac{ 2 \beta^{-1} q^3 r^{\mu - 1} }{ \left( r^\nu + q^\nu \right)^{\frac{\mu}{ \nu}} },
\end{align}
arises as an exact solution of four-dimensional general relativity coupled to nonlinear electrodynamics (NED) with action
\begin{align}
I = \frac{1}{ 16 \pi G } \int d^4 x \sqrt{-g} \left( R - {\cal L} \left( {\cal F} \right) \right),
\qquad
{\cal L} \left( {\cal F} \right) = \frac{4 \mu}{\beta} \frac{ \left( \beta {\cal F} \right)^{\frac{\nu + 3}{4}} }{ \left( 1 + \left( \beta {\cal F} \right) \right)^{\frac{\mu + \nu}{\nu}} },
\qquad
{\cal F} = F_{\mu \nu} F^{\mu \nu},
\end{align}
where $\mu$, $\nu$, and $\beta$ are free parameters of the theory.

The solution~(\ref{eq:fw}) is obtained for a purely magnetic configuration,
\begin{align}
A = \frac{q^2}{\sqrt{2\beta}} \cos\theta\, d\phi,
\end{align}
for which the ADM mass and magnetic charge are given by $M_{\rm ADM}=M+q^3\beta^{-1}$ and $Q_{\rm m}=q^2/\sqrt{2\beta}$, respectively.
The metric~(\ref{eq:fw}) is free from curvature singularities at the center $r=0$ provided that $M=0$ and $\mu\ge 3$, in which case it reduces to
\begin{align}
\label{eq:fw_regular}
ds^2 = - f(r)\,dt^2 + \frac{dr^2}{ f(r) } + r^2 d \Omega_2^2,
\qquad
f(r) = 1 -  \frac{ 2 \beta^{-1} q^3 r^{\mu - 1} }{ \left( r^\nu + q^\nu \right)^{\frac{\mu}{ \nu}} }.
\end{align}
The Bardeen- and Hayward-type regular black holes are recovered by choosing $(\mu,\nu)=(3,2)$ and $(\mu,\nu)=(3,3)$, respectively~\cite{Bardeen:1968aa,Hayward:2005gi}.
Depending on the charge parameter $q$, the regular solution describes three possible spacetimes
~\cite{Isomura:2023oqf}:
\begin{enumerate}
\item a black hole with two horizons: $0<q<q_{\rm ex}$,
\item an extremal black hole with a degenerate horizon: $q=q_{\rm ex}$,
\item an overcharged but regular horizonless spacetime: $q>q_{\rm ex}$,
\end{enumerate}
where the extremal charge is determined by
\begin{align}
q_{\rm ex} = \sqrt{ \frac{ \beta \mu^{\frac{\mu}{ \nu}} }{ 2 (\mu - 1)^{\frac{\mu - 1}{ \nu }} } }.
\end{align}




\section{Fan--Wang black holes in higher dimensions}\label{sec:hdfw}

Now, we study the higher-dimensional generalization of the Fan--Wang type metric in four dimensions~\cite{Fan:2016hvf}, which was briefly reviewed in the previous section.
Below, we demonstrate that the following higher dimensional Fan--Wang metric becomes a solution for vacuum QTG under the appropriate coupling constants:
\begin{align}
\label{eq:hdfw}
ds^2 =& - f(r) dt^2 + \frac{dr^2}{f(r)} + r^2 d \Omega_{D-2}^2,
\quad
f(r) = 1 - \frac{ m r^{\frac{(\mu - 3)(D-1)}{3} + 2} }{ \left( r^{\frac{\nu (D-1)}{3} } + \alpha^{\frac{\nu}{3}} m^{\frac{\nu}{3}} \right)^{\frac{\mu}{\nu}}}.
\end{align}
This metric reduces to the regular Fan--Wang metric~(\ref{eq:fw_regular}) with $\beta = 2 \alpha $ and $ q = \sqrt[3]{\alpha m} $ for $D=4$. If we choose $(\mu , \nu) = (3,3)$, 
it also reproduces the higher dimensional Hayward black hole (\ref{eq:Hayward}) in Ref.~\cite{Bueno:2024dgm}.
\medskip

\subsection{Derivation from QTG}
First, we show the metric~(\ref{eq:hdfw}) is the solution for QTG by finding the expression for the coupling constants $\alpha_n$.
If the metric~(\ref{eq:hdfw}) satisfies the field equations of QTG, 
eqs.~(\ref{eq:psidef}) and (\ref{eq:hsolution}) must hold for
\begin{align}
\label{eq:psi_hdfw}
\psi :=\frac{1-f(r)}{r^2}= \frac{ m r^{\frac{(\mu - 3)(D-1)}{3}} }{ \left( r^{ \frac{\nu (D-1)}{3} } + \alpha^{\frac{\nu}{3}} m^{\frac{\nu}{3}} \right)^{\frac{\mu}{ \nu}} },
\end{align}
where we identify the mass parameter $m$ in the solution with that in eq.~(\ref{eq:hsolution}).
Substituting eq.~(\ref{eq:hsolution}) into eq.~(\ref{eq:psi_hdfw}), we obtain
\begin{align}
\label{eq:psi_h}
\psi = \frac{ h }{ \left( 1 + \alpha^{\frac{\nu}{3}} h^{\frac{\nu}{3}} \right)^{\frac{\mu}{ \nu}} }
=: \Psi (h),
\end{align}
where we assume the analyticity of $\Psi(h)$ at $h=0$ by requiring the parameter $\nu$ to be a positive multiple of three:
\begin{align}
\nu = \left\{ 3 \bar{\nu}\, |\, \bar{\nu} \in \mathbb{N} \right\}.
\end{align}
Therefore, the expression for $h(\psi)$ in eq.~(\ref{eq:psidef}) is implicitly given as the inverse function of $\Psi(h)$, and the curvature coefficients $\alpha_n$ are determined by finding the series expansion of $h(\psi)$ in the power of $\psi$.
The series expansion of an inverse function is formally written by using the Lagrange inversion formula~\cite{Gessel2016,Surya2023}\footnote{
For $x = f(y)$ with $x_0 = f(y_0)$ and $f'(y_0) \neq 0$ then, the series expansion of $y= f^{-1}(x)$ around $x_0$ is given by
\begin{align}
y = y_0 + \sum_{n=1}^\infty g_n  \left( x - x_0 \right)^n ,\quad 
g_n := \frac{1}{n!} \lim_{y \to y_0} \frac{d^{n-1}}{dy^{n-1}} \left( \frac{y-y_0}{ f(y) - x_0 } \right)^n.
\end{align}
}
\begin{align}
\label{eq:hLB}
h(\psi) = \psi + \sum_{n=2}^\infty \alpha_n \psi^n,\quad 
\alpha_n =& \frac{1}{n!} \lim_{h \to 0} \frac{d^{n-1}}{dh^{n-1}} \left( \frac{h}{ \Psi(h) } \right)^n.
\end{align}
To obtain the explicit form of $\alpha_n$, we further expand a part of the expression as
\begin{align}
 \left( \frac{h}{ \Psi(h) } \right)^n=(1+(\alpha h)^{\frac{\nu}{3}})^{\frac{n\mu}{\nu}} =
 \sum_{l=0}^{\infty} G_l (h) , \quad G_l (h) =& \frac{ \Gamma ( \frac{n \mu}{ \nu} +1) }{ \Gamma ( l + 1 ) \Gamma ( \frac{n \mu}{ \nu} - l + 1) } \alpha^{\frac{l \nu}{3}} h^{ \frac{l \nu}{ 3} }.
\end{align}
One can calculate the coefficients as follows
\begin{align}\label{eq:alpha_n_LB_calc}
\alpha_n = \frac{1}{n!} \lim_{h \to 0} \frac{ d^{n-1} }{ dh^{n-1} } \sum_{l =0}^{\infty} G_l  (h)
= \frac{1}{n!}  \sum_{l =0}^\infty \lim_{h \to 0} \frac{d^{n-1}}{dh^{n-1}} G_l  (h),
\end{align}
where we exchanged the infinite sum and the limit in the second equality using the fact that the series expansion $\sum_{l =0}^\infty G_l (h)$ uniformly converges to $(1+(\alpha h)^{\bar{\nu}})^{n\mu/(3\bar{\nu})}$ around a neighborhood of $h=0$.
Since  $G_l(h) \propto h^{l \bar \nu}$ and $l\bar{\nu}$ is a nonnegative integer,
one can easily see that $ \frac{d^{n-1}}{dh^{n-1}} G_l  (h)$ vanishes unless  $ l  \bar{\nu} \geq n-1 $  
\begin{align}
\frac{d^{n-1}}{dh^{n-1}} G_l  (h) =
\begin{cases}
\displaystyle
\frac{ \Gamma \left( \frac{n \mu}{3 \bar{\nu}} +1 \right) }{ \Gamma ( l  + 1 ) \Gamma \left( \frac{n \mu}{3 \bar{\nu}} - l  + 1 \right) } \alpha^{l  \bar{\nu}} \frac{ \left( l  \bar{\nu} \right) ! }{ \left( l  \bar{\nu} - n+1 \right) ! } h^{ l  \bar{\nu} - n+1 } & \left(l  \bar{\nu} \geq (n-1)  \right) \\
0 & \left( l  \bar{\nu} < (n-1) \right)
\end{cases}.
\end{align}

This means that only the term with $l  = (n-1)/\bar{\nu}$ contributes in eq.~(\ref{eq:alpha_n_LB_calc}) at the limit $h \to 0$. 
Thus, the only nonzero coefficients are those with $n = l  \bar{\nu} +1$ for $(l  =0,1,2,\ldots)$
\begin{align}\label{eq:coupling}
\alpha_{l  \bar{\nu}+1} = \frac{1}{l  \bar{\nu}+1} \frac{ \Gamma ( \frac{l  \mu}{ 3}+ \frac{\mu}{ 3 \bar{\nu }} +1) }{ \Gamma ( l + 1 ) \Gamma ( \left(\frac{ \mu }{ 3 }-1\right)l  + \frac{  \mu }{ 3 \bar{\nu} } + 1) } \alpha^{l  \bar{\nu} } \quad {\rm for} \quad  l  =0,1,2,\ldots.
\end{align}




\subsection{Regularity at the center}

\label{ssec:reg_cneter}

We now analyze the regularity at the center $r=0$.
The requirements of regularity at the center and consistency with the QTG solution uniquely fix one of the parameters, $\mu$, to $\mu=3$.
Indeed, we show that consistency with the behavior of the solution~(\ref{eq:hsolution}) near $r=0$ imposes $\mu \leq 3$, whereas regularity of the metric function $f(r)$ at $r=0$ requires $\mu \geq 3$.

\medskip
First, one can easily find $\mu \leq 3$ from the consistency in the formalism as follows.
If we assume $\mu > 3$, eq.~(\ref{eq:psi_hdfw}) leads to $\psi(r) \to 0$ at $r \to 0$. 
On the other hand, the result of the field equation~(\ref{eq:hsolution}) implies that $h(\psi) \to \infty$ at $r\to0$.
These behaviors at $r\to 0$ contradict the analyticity of $h(\psi)$ at $\psi =0$ in eq.~(\ref{eq:hLB}).
Hence, we must impose $\mu \leq 3$ for the consistency.

\medskip
Next, we examine the spacetime geometry in the vicinity of the center $r=0$ in more detail.
Near $r=0$, the metric function $f(r)$ appearing in Eq.~(\ref{eq:hdfw}) admits the following asymptotic expansion:
\begin{align}
f(r) \simeq 1 - \alpha^{-\frac{\mu}{3}} m^{1-\frac{\mu}{3}}
\, r^{\frac{(\mu-3)(D-1)}{3} + 2}.
\end{align}
The regularity condition at the center, given in Eq.~(\ref{eq:regularity_f}),
requires that the leading correction to one be at least of order $r^2$.
This condition is satisfied if
\begin{align}
\frac{(\mu-3)(D-1)}{3} + 2 \geq 2,
\end{align}
which leads to the inequality $\mu \geq 3$.

\medskip
Thus, the only choice satisfying both the consistency and regularity is $\mu=3$.
Note that, while the first constraint comes from the theoretical feature of QTG, the latter is also required in the four-dimensional solution~\cite{Fan:2016hvf} for the same reasoning.
Below, we take a closer look at the center geometry in the $\mu=3$ case.

\medskip
By setting $\mu=3$, the metric~(\ref{eq:hdfw}) reduces to

\begin{align}
\label{eq:hdfw_qtg}
ds^2 =& - f(r) dt^2 + \frac{dr^2}{f(r)} + r^2 d \Omega_{D-2}^2,
\quad
f(r) = 1 - \frac{ m r^2 }{ \left( r^{ \bar{\nu} (D-1) } + \alpha^{\bar{\nu}} m^{\bar{\nu}} \right)^{ \frac{1}{ \bar{\nu}}} }
\quad
\left( \bar{\nu} \in \mathbb{N} \right).
\end{align}
Note that, for $\mu=3$, we can explicitly solve eq. (\ref{eq:psi_h}) with respect to $h$ as
\begin{align}
\label{eq:h(psi)_closed}
h(\psi) = \frac{ \psi }{ \left( 1 - \alpha^{\bar{\nu}} \psi^{\bar{\nu}} \right)^{\frac{1}{\bar{\nu}}} }.
\end{align}
Near the center, the metric admits a de Sitter core
\begin{align}
ds^2 \simeq - \left( 1 - \frac{ r^2 }{ \alpha } \right) dt^2 + \left( 1 + \frac{ r^2 }{ \alpha } \right) dr^2 + r^2 d \Omega_{D-2}^2.
\end{align}
Moreover, to examine the differentiability of the spacetime at the center, we introduce the Cartesian coordinate system $ (t,\bm{x}) = (t , x^1 , x^2 , \cdots , x^{D-1})$, in which the metric (\ref{eq:hdfw_qtg}) is expressed as
\begin{align}
ds^2 
=& - f( \sqrt{ \bm{x} \cdot \bm{x} } ) dt^2 + \left\{ \delta_{ij} + \left( \frac{1}{ f(\sqrt{ \bm{x} \cdot \bm{x} }) } - 1 \right) \frac{x_i x_j}{ \bm{x} \cdot \bm{x} } \right\} dx^i dx^j .
\end{align}
This can be expanded around the center as follows
\begin{align}
ds^2& \simeq - \left(1- \frac{\bm{x} \cdot \bm{x} }{\alpha} + \frac{1}{\alpha} \frac{ (\bm{x}\cdot \bm{x})^{\frac{\bar{\nu}(D-1)}{2}+1} }{\bar{\nu}\alpha^{\bar{\nu}} m^{\bar{\nu}}}\right)dt^2
+\left( \delta_{ij} + \frac{ x_i x_j }{ \alpha } \left( 1 + \frac{ \bm{x} \cdot \bm{x} }{\alpha} - \frac{ \left( \bm{x} \cdot \bm{x} \right)^{ \frac{\bar{\nu} (D-1)}{ 2} } }{ \bar{\nu} \alpha^{ \bar{\nu} } m^{ \bar{\nu} } } \right)\right) dx^i dx^j.
\end{align}
Therefore, when $D$ is odd or when $\bar{\nu}$ is even, the solution is $C^\infty$ at the center,
and when $D$ is even and $\bar{\nu}$ is odd, the solution is at least $C^{ \bar{\nu} (D-1) + 2 }$ as summarized in Table \ref{tab:diff_center}.

\begin{table}[t]
\begin{align*}
\begin{array}{|c|c|c|}
\hline
& \bar{\nu} ~ \mbox{odd} & \bar{\nu} ~ \mbox{even} \\
\hline
D ~ \mbox{odd} & C^\infty & C^\infty \\
\hline
D ~ \mbox{even} & C^{ \bar{\nu} (D-1) + 2 } & C^\infty \\
\hline
\end{array}
\end{align*}
\caption{Differentiability at the center.}
\label{tab:diff_center}
\end{table}




\subsection{Asymptotic flatness and thermodynamics}

\label{ssec:flat}

The asymptotic flatness of the regular solution~(\ref{eq:hdfw_qtg}) follows from the expansion at the limit $r\to\infty$,
\begin{align}
ds^2 \simeq - \left( 1 - \frac{m}{ r^{D-3} } \right) dt^2
+ \left( 1 + \frac{m}{ r^{D-3} } \right) dr^2
+ r^2 d \Omega_{D-2}^2,
\end{align}
Since in QTG,  the asymptotic charges can be introduced as in general relativity~\cite{Bueno:2024dgm}, one can read off the ADM mass from the above behavior as
\begin{align}
M =& \frac{ (D-2) \pi^{\frac{D-3}{2}} m }{ 8 G\, \Gamma \left( \frac{D-1}{2} \right) }. 
\end{align}
This is equivalent to the general expression in Ref.~\cite{Bueno:2024dgm} through eq.~(\ref{eq:hsolution}),
\begin{align}
M =& \frac{ (D-2) \pi^{\frac{D-3}{2}} r_{+}^{D-1} }{ 8 G \Gamma \left( \frac{D-1}{2} \right) } h (\psi_{+}),
\end{align}
where we wrote the position of the event horizon as $r_+$ defined as the zero of $f(r_+)=0$, and $\psi_+:=\psi(r_+)=r_+^{-2}$.
Following Ref.~\cite{Bueno:2024dgm}, one can also calculate
the Wald entropy $S$ and Hawking temperature $T$ at the event horizon as
\begin{align}
\nonumber
S =& - \frac{ (D-2) \pi^{\frac{D-1}{2} } }{4G \Gamma \left( \frac{D-1}{2} \right)} \int \frac{ h' (\psi_{+}) }{ \psi_{+}^{\frac{D}{2}} } d \psi_{+} \\
=& \frac{ \pi^{\frac{D-1}{2} } r_{+}^{D-2} }{2G \Gamma \left( \frac{D-1}{2} \right)} \,{}_2 F_1 \left( 1 + \frac{1}{\bar{\nu}} , - \frac{ D - 2 }{ 2 \bar{\nu} } ; 1 - \frac{ D - 2 }{ 2 \bar{\nu} } ; \left( \frac{ \alpha }{r_{+}^2} \right)^{\bar{\nu}}  \right) , \\
T =& \frac{1}{ 4 \pi r_{+} } \left[ \frac{ (D-1) r_{+}^2 h (\psi_{+}) }{ h' (\psi_{+}) } - 2 \right]
= \frac{1}{ 4 \pi r_{+} } \left[ (D-1) \left( 1 - \left( \frac{ \alpha }{ r_{+}^2 } \right)^{ \bar{\nu} } \right) - 2 \right],
\end{align}
where the expression of $h(\psi)$ is given by eq.~(\ref{eq:h(psi)_closed}) and $ {}_2 F_1 \left( a , b ; c ; z \right) $ is the hypergeometric function defined as $ \displaystyle {}_2 F_1 \left( a , b ; c ; z \right) = \frac{ \Gamma (c) }{ \Gamma (b) \Gamma (c-b) } \int_0^1 t^{b-1} (1-t)^{c-b-1} (1 - tz)^{-a} dt $.
One can easily confirm the above quantities satisfy the first law of thermodynamics
\begin{align}
dM =& T d S.
\end{align}




\subsection{Horizons}

\label{ssec:horizons}
Here, we study the properties of the event horizon.
The locations of the horizons in the metric~(\ref{eq:hdfw_qtg})
are determined by the roots of
\begin{align}
f(r) = 0.
\end{align}
It is easy to check that the spacetime admits two horizons at most as follows.
At the center $r=0$ and infinity $r=\infty$, $f(r)$ takes positive values:
\begin{align}
\lim_{r \to 0} f(r) = \lim_{r \to \infty} f(r) = 1.
\end{align}
In addition, the first derivative
\begin{align}
f'(r) = 
\frac{ m r \left[ \left( D - 3 \right) r^{ \bar{\nu} \left( D - 1 \right)  } - 2 \alpha^{ \bar{\nu} } m^{ \bar{\nu} } \right] }
{ \left( r^{ \bar{\nu} \left( D - 1 \right) } + \alpha^{ \bar{\nu} } m^{ \bar{\nu} } \right)^{ \frac{1}{ \bar{\nu}} + 1 } }
\end{align}
has a single zero-point $r_0$ in the range $0 < r < \infty$:
\begin{align}
f'(r)
\begin{cases}
< 0 & (0 < r < r_0) \\
= 0 & (r = r_0) \\
> 0 & (r_0 < r < \infty)
\end{cases},
\quad
r_0 = \sqrt[D-1]{ \left( \frac{ 2 }{ D - 3 } \right)^\frac{1}{\bar{\nu} } \alpha m }.
\end{align}
Therefore, the spacetime admits at most two horizons.
We denote $r_+$ and $r_-$ the radii of the outer and inner horizons, respectively:
\begin{align}
f(r_\pm) = 0
\quad
(r_- \leq r_+).
\end{align}
The two horizons coincide in the extremal limit, $r_- = r_+$, which is equivalent to
\begin{align}
f (r_0) = 0.
\end{align}

This can be solved as
\begin{align}
m =& 
\left( \frac{ 2 }{ D - 3 } \right)^{ \frac{ D - 3 }{ 2 \bar{\nu} } }
\left( \frac{ D - 1 }{  2 } \right)^{ \frac{ D - 1 }{ 2 \bar{\nu} } }
\alpha^{ \frac{ D - 3 }{ 2 } }
=: m_{\rm ex}, 
\\
r_0 =& 
\left( \frac{ D-1 }{ D - 3 } \right)^{ \frac{ 1 }{ 2 \bar{\nu} } }
\alpha^{ \frac{ 1 }{ 2 } }
=: r_{\rm ex}.
\end{align}
Hence we have three distinct cases depending on $m$:
\begin{enumerate}
\item a horizonless spacetime : $ m < m_{\rm ex} $.
\item a black hole with a degenerate horizon : $ m = m_{\rm ex} $.
\item a black hole with outer and inner horizons : $ m > m_{\rm ex} $.
\end{enumerate}

\medskip


 In the nondegenerate case, the regularity of the outer and inner horizons $(r=r_{\pm})$, including the corresponding white hole horizons, can be examined by introducing the Eddington--Finkelstein coordinates $v$ and $u$:
\begin{align}
\label{eq:efcood}
v = t + \int \frac{dr}{f(r)}, \qquad 
u = t - \int \frac{dr}{f(r)} .
\end{align}
In terms of the outgoing coordinate $v$, the metric becomes
\begin{align}
ds^2 = -f(r)\,dv^2 + 2\,dv\,dr + r^2 d\Omega_{D-2}^2,
\end{align}
while in terms of the ingoing coordinate $u$, it can be written as
\begin{align}
ds^2 = -f(r)\,du^2 - 2\,du\,dr + r^2 d\Omega_{D-2}^2 .
\end{align}
Since the metric function $f(r)$ is manifestly $C^{\infty}$ on the null hypersurfaces $r=r_{\pm}$, 
the metric itself is also $C^{\infty}$ there. Consequently, no curvature singularities arise at either the outer or inner horizons, including the white hole horizon.

\medskip
In the extremal case, the metric behaves around $r=r_{\rm ex}$  as 
\begin{align}
  ds^2 \simeq - A (r-r_{\rm ex})^2 dt^2 + \frac{dr^2}{A(r-r_{\rm ex})^2} + r_{\rm ex}^2 d\Omega^2,
\end{align}
where
\begin{align}
 A := \frac{1}{2} f''(r_{\rm ex}) =  \frac{\bar{\nu} (D-3)}{\alpha} \left(\frac{D-3}{D-1}\right)^\frac{1}{\bar{\nu}}.
\end{align}
Similarly, in terms of the new coordinates $(v,X)$ by
\begin{align}
  r = r_{\rm ex} + A^{-\frac{1}{2}} X,\quad  t =  v +X^{-1},
\end{align}
the metric near the horizon $r= r_{\rm ex}$ behaves as
\begin{align}
  ds^2 \simeq - X^2 dv^2 +2 dv dX+ r_{\rm ex}^2 d\Omega^2,
\end{align}
which is $C^{\infty}$ on the horizon and hence implies the absence of curvature singularities throughout the spacetime.




\subsection{Negative mass}

\label{ssec:mass}

In the analysis so far, we have implicitly assumed the positivity of the mass parameter $m>0$.
Here we discuss the properties of the solutions with negative mass parameter $m<0$.
For $m<0$, the metric function takes the form
\begin{align}
f(r) = 1 + \frac{ \left| m \right| r^2 }{ \left( r^{ \bar{\nu} (D-1) } + \left( -1 \right)^{\bar{\nu}} \alpha^{\bar{\nu}} \left| m \right|^{\bar{\nu}} \right)^{\frac{1}{\bar{\nu}}} },
\quad
\psi = - \frac{ \left| m \right| }{ \left( r^{ \bar{\nu} (D-1) } + \left( -1 \right)^{ \bar{\nu} } \alpha^{ \bar{\nu} } \left| m \right|^{ \bar{\nu} } \right)^{\frac{1}{ \bar{\nu}}} } .\label{eq:fr-negativemass}
\end{align}

\medskip

If $\bar{\nu}$ is even, 
the sign inside the parenthesis in the denominator of $\psi$ remains positive for all $r \geq 0$,
leading to both the positive definiteness and finiteness of $f(r)$ for $r \geq 0$.
Such  a spacetime describes a horizonless spacetime free from curvature singularity, 
but admits negative mass.

\medskip

If $\bar{\nu}$ is odd, the expression inside the parentheses in the denominator
of $\psi$ vanishes at the radius
\begin{align}
r = \left( \alpha |m| \right)^{\frac{1}{D-1}} =: r_{\rm s}.
\end{align}
At this radius, the function $\psi$ becomes non-analytic and diverges, and the
same is true for its derivatives $\psi'$ and $\psi''$.
As a consequence, the curvature invariants
(\ref{eq:ricciscal})--(\ref{eq:kretschmann}) also diverge at $r=r_{\rm s}$.
For instance, the Kretschmann scalar behaves as
\begin{align}
R_{\mu\nu\rho\sigma}R^{\mu\nu\rho\sigma} \propto (r-r_{\rm s})^{-4-\frac{2}{\bar{\nu}}}
\end{align}
Therefore, the spacetime contains naked curvature singularities located on the surface 
$r=r_{\rm s}$.




\section{Summary}

\label{sec:summary}

In this work, we have investigated a higher-dimensional generalization of the Fan--Wang class of regular black holes within the framework of quasi-topological gravity (QTG) in spacetime dimensions $D>4$. 
Unlike the original four-dimensional Fan--Wang solution, which arises in Einstein gravity coupled to nonlinear electrodynamics, the present solution is obtained purely as a vacuum solution of QTG by appropriately tuning the coupling constants of an infinite tower of higher-curvature corrections. 
We have derived the explicit form of the coupling constants required for the metric to satisfy the QTG field equations, thereby establishing a concrete realization of Fan--Wang--type regular geometries in higher dimensions without introducing matter fields.
The spacetime is characterized by the mass parameter $m$ and two additional parameters $(\mu,\nu)$, which are encoded in the structure of the higher-curvature couplings. 
We have shown that consistency of the formalism together with regularity at the center imposes strong constraints on these parameters. 
In particular, analyticity of the inverse function $h(\psi)$ requires $\nu = 3\bar{\nu}$ with $\bar{\nu} \in \mathbb{N}$, while the combined requirements of regularity and consistency uniquely fix $\mu = 3$. 
This restriction is stronger than in the four-dimensional Einstein-NED case, and highlights a structural constraint intrinsic to QTG.

\medskip
Under these conditions, the solution possesses a regular center with a de Sitter core. 
We have analyzed the differentiability properties of the spacetime at $r=0$, showing that the metric is $C^\infty$ in odd dimensions or for even $\bar{\nu}$, while in even dimensions with odd $\bar{\nu}$ the differentiability is finite but still sufficiently high. 
This confirms that the curvature invariants remain finite and that the central region is genuinely nonsingular.
We have also demonstrated that the solution is asymptotically flat. 
The ADM mass can be read off from the asymptotic expansion and agrees with the general expression obtained in QTG. 
Furthermore, the Wald entropy and Hawking temperature were computed explicitly, and we have verified that they satisfy the first law of black hole thermodynamics. 
These results confirm the thermodynamic consistency of the higher-dimensional Fan--Wang solution within the QTG framework.

\medskip
For positive mass parameter $m>0$, the spacetime exhibits a structure qualitatively similar to that of charged black holes: depending on the value of $m$, the geometry admits either two distinct horizons, a single degenerate (extremal) horizon, or no horizon. 
In the non-degenerate case, we introduced Eddington-Finkelstein coordinates and confirmed that the horizons are regular null hypersurfaces rather than curvature singularities. 
The extremal configuration is characterized by the coincidence of the two horizons and is determined explicitly by analytic expressions for $m_{\rm ex}$ and $r_{\rm ex}$.
For negative mass, $m<0$, the global structure depends sensitively on the parity of $\bar{\nu}$. When $\bar{\nu}$ is odd, a naked curvature singularity inevitably develops at a finite radius, so that the spacetime becomes physically unacceptable from the asymptotic region. In contrast, when $\bar{\nu}$ is even, the geometry remains completely regular and horizonless despite the negative mass. This peculiar behavior indicates that quasi-topological gravity does not necessarily obey a positive-mass property in the same sense as Einstein gravity, raising an important conceptual issue that merits further investigation.

\medskip
Throughout this paper, we have restricted attention to static and spherically symmetric configurations. 
It would be natural to explore whether rotating generalizations of the present solution exist within QTG, and whether similar regularization mechanisms persist in the presence of angular momentum. 
Another interesting direction is to investigate the stability of these regular black holes under linear perturbations. 
Finally, it would be worthwhile to clarify the precise relation between the four-dimensional NED realization of Fan--Wang black holes and their vacuum realization in non-polynomial QTG, thereby deepening our understanding of how higher-curvature corrections mimic effective matter sources.
We expect that the present analysis provides further insight into the role of infinite higher-curvature corrections in resolving classical singularities and contributes to the broader program of constructing consistent and physically viable regular black hole solutions in modified theories of gravity.




\acknowledgments
RS was supported by JSPS KAKENHI Grant Number JP24K07028. 
ST was supported by JSPS KAKENHI Grant Number JP21K03560.





\begin{thebibliography}{99}








\bibitem{LIGOScientific:2016aoc}
B.~P.~Abbott \textit{et al.} [LIGO Scientific and Virgo],
``Observation of Gravitational Waves from a Binary Black Hole Merger,''
Phys. Rev. Lett. \textbf{116}, no.6, 061102 (2016)
[arXiv:1602.03837 [gr-qc]].



\bibitem{EventHorizonTelescope:2019dse}
K.~Akiyama \textit{et al.} [Event Horizon Telescope],
``First M87 Event Horizon Telescope Results. I. The Shadow of the Supermassive Black Hole,''
Astrophys. J. Lett. \textbf{875}, L1 (2019)
[arXiv:1906.11238 [astro-ph.GA]].



\bibitem{Penrose:1964wq}
R.~Penrose,
``Gravitational collapse and space-time singularities,''
Phys. Rev. Lett. \textbf{14}, 57-59 (1965).






\bibitem{Bardeen:1968aa}
J .M . Bardeen, ``Non-singular general-relativistic gravitational collapse.'' 
in Proceedings of International Conference GR5 (Tbilisi, USSR, 1968) p. 174.


\bibitem{Dymnikova:1992ux}
I.~Dymnikova,
``Vacuum nonsingular black hole,''
Gen. Rel. Grav. \textbf{24}, 235-242 (1992)



\bibitem{Hayward:2005gi}
S.~A.~Hayward,
``Formation and evaporation of regular black holes,''
Phys. Rev. Lett. \textbf{96}, 031103 (2006)
[arXiv:gr-qc/0506126 [gr-qc]].

\bibitem{Ansoldi:2008jw}
S.~Ansoldi,
``Spherical black holes with regular center: A Review of existing models including a recent realization with Gaussian sources,''
[arXiv:0802.0330 [gr-qc]].

\bibitem{Ayon-Beato:1998hmi}
E.~Ay{\'o}n-Beato and A.~Garc{\'i}a,
``Regular black hole in general relativity coupled to nonlinear electrodynamics,''
Phys. Rev. Lett. \textbf{80}, 5056-5059 (1998)
[arXiv:gr-qc/9911046 [gr-qc]].



\bibitem{Ayon-Beato:2000mjt}
E.~Ay{\'o}n-Beato and A.~Garc{\'i}a,
``The Bardeen model as a nonlinear magnetic monopole,''
Phys. Lett. B \textbf{493}, 149-152 (2000)
[arXiv:gr-qc/0009077 [gr-qc]].






\bibitem{Fan:2016hvf}
Z.~Y.~Fan and X.~Wang,
``Construction of Regular Black Holes in General Relativity,''
Phys. Rev. D \textbf{94}, no.12, 124027 (2016)
[arXiv:1610.02636 [gr-qc]].



\bibitem{Dymnikova:2015hka}
I.~Dymnikova and E.~Galaktionov,
``Regular rotating electrically charged black holes and solitons in non-linear electrodynamics minimally coupled to gravity,''
Class. Quant. Grav. \textbf{32}, no.16, 165015 (2015)
[arXiv:1510.01353 [gr-qc]].



\bibitem{Takeuchi:2016nrj}
S.~Takeuchi,
``Kerr/CFT correspondence in a 4D extremal rotating regular black hole with a non-linear magnetic monopole,''
Nucl. Phys. B \textbf{921}, 375-393 (2017)
[arXiv:1609.09427 [hep-th]].



\bibitem{Kruglov:2017fck}
S.~I.~Kruglov,
``Black hole as a magnetic monopole within exponential nonlinear electrodynamics,''
Annals Phys. \textbf{378}, 59-70 (2017)
[arXiv:1703.02029 [gr-qc]].



\bibitem{Maeda:2021jdc}
H.~Maeda,
``Quest for realistic non-singular black-hole geometries: regular-center type,''
JHEP \textbf{11}, 108 (2022)
[arXiv:2107.04791 [gr-qc]].



\bibitem{Bokulic:2022cyk}
A.~Bokuli\'c, I.~Smoli\'c and T.~Juri\'c,
``Constraints on singularity resolution by nonlinear electrodynamics,''
Phys. Rev. D \textbf{106}, no.6, 064020 (2022)
[arXiv:2206.07064 [gr-qc]].



\bibitem{Bronnikov:2022ofk}
K.~A.~Bronnikov,
``Regular black holes sourced by nonlinear electrodynamics,''
[arXiv:2211.00743 [gr-qc]].



\bibitem{Frolov:2016pav}
V.~P.~Frolov,
``Notes on nonsingular models of black holes,''
Phys. Rev. D \textbf{94}, no.10, 104056 (2016)
[arXiv:1609.01758 [gr-qc]].

\bibitem{Oliva:2010eb}
J.~Oliva and S.~Ray,
``A new cubic theory of gravity in five dimensions: Black hole, Birkhoff's theorem and C-function,''
Class. Quant. Grav. \textbf{27}, 225002 (2010)
[arXiv:1003.4773 [gr-qc]].

\bibitem{Myers:2010ru}
R.~C.~Myers and B.~Robinson,
``Black Holes in Quasi-topological Gravity,''
JHEP \textbf{08}, 067 (2010)
[arXiv:1003.5357 [gr-qc]].



\bibitem{Dehghani:2011vu}
M.~H.~Dehghani, A.~Bazrafshan, R.~B.~Mann, M.~R.~Mehdizadeh, M.~Ghanaatian and M.~H.~Vahidinia,
``Black Holes in Quartic Quasitopological Gravity,''
Phys. Rev. D \textbf{85}, 104009 (2012)
[arXiv:1109.4708 [hep-th]].


\bibitem{Cisterna:2017umf}
A.~Cisterna, L.~Guajardo, M.~Hassaine and J.~Oliva,
``Quintic quasi-topological gravity,''
JHEP \textbf{04}, 066 (2017)
[arXiv:1702.04676 [hep-th]].




\bibitem{Bueno:2016xff}
P.~Bueno and P.~A.~Cano,
``Einsteinian cubic gravity,''
Phys. Rev. D \textbf{94}, no.10, 104005 (2016)
[arXiv:1607.06463 [hep-th]].

\bibitem{Hennigar:2017ego}
R.~A.~Hennigar, D.~Kubiz{\v{n}}{\'a}k and R.~B.~Mann,
``Generalized quasitopological gravity,''
Phys. Rev. D \textbf{95}, no.10, 104042 (2017)
[arXiv:1703.01631 [hep-th]].
\bibitem{Bueno:2019ycr}
P.~Bueno, P.~A.~Cano and R.~A.~Hennigar,
``(Generalized) quasi-topological gravities at all orders,''
Class. Quant. Grav. \textbf{37}, no.1, 015002 (2020)
[arXiv:1909.07983 [hep-th]].

\bibitem{Aguayo:2025xfi}
M.~Aguayo, L.~Gajardo, N.~Grandi, J.~Moreno, J.~Oliva and M.~Reyes,
``Holographic explorations of regular black holes in pure gravity,''
JHEP \textbf{09}, 030 (2025)
[arXiv:2505.11736 [hep-th]].


\bibitem{Bueno:2024dgm}
P.~Bueno, P.~A.~Cano and R.~A.~Hennigar,
``Regular black holes from pure gravity,''
Phys. Lett. B \textbf{861}, 139260 (2025)
[arXiv:2403.04827 [gr-qc]].




\bibitem{Fernandes:2025fnz}
P.~G.~S.~Fernandes,
``Singularity resolution and inflation from an infinite tower of regularized curvature corrections,''
Phys. Rev. D \textbf{112}, no.8, 8 (2025)
[arXiv:2504.07692 [gr-qc]].

\bibitem{Hao:2025utc}
C.~H.~Hao, J.~Jing and J.~Wang,
``Charged Regular Black Holes From Quasi-topological Gravities in $D\ge 5$,''
[arXiv:2512.04604 [gr-qc]].





\bibitem{Bueno:2025zaj}
P.~Bueno, P.~A.~Cano, R.~A.~Hennigar and {\'A}.~J.~Murcia,
``Regular black hole formation in four-dimensional non-polynomial gravities,''
[arXiv:2509.19016 [gr-qc]].

\bibitem{Isomura:2023oqf}
K.~Isomura, R.~Suzuki and S.~Tomizawa,
``Particle motions around regular black holes,''
Phys. Rev. D \textbf{107}, no.8, 084003 (2023)
[arXiv:2301.10465 [gr-qc]].





\bibitem{Gessel2016}
Ira M. Gessel, 
``Lagrange inversion,''
Journal of Combinatorial Theory, Series A, \textbf{144}, 212-249 (2016)
[arXiv:1609.05988 [math.CO]]



\bibitem{Surya2023}
E. Surya and L. Warnke,
``Lagrange Inversion Formula by Induction,''
The American Mathematical Monthly, \textbf{130} (10), 944-948 (2023)
[arXiv:2305.17576 [math.CO]]





\end{thebibliography}
\end{document}